\documentclass[12pt,a4paper]{article}
\usepackage{amsmath,amssymb}
\usepackage{cite}
\usepackage[utf8]{inputenc}
\usepackage{graphicx}
\usepackage{dcolumn}
\usepackage{bm}
\usepackage{hyperref}

\newcommand{\bra}[1]{\langle #1 |}
\newcommand{\ket}[1]{| #1 \rangle}

\title{Bohmian Field Theory on a Shape Dynamics Background and Unruh Effect}

\author{Furkan Semih Dündar\footnote{email: furkan.dundar@boun.edu.tr}$\,^{,a}$ and Metin Arık\footnote{email: metin.arik@boun.edu.tr}$\,^{,a}$\\ 
\footnotesize $^a$\textit{Physics Department, Boğaziçi University, 34342, Istanbul, Turkey}
}

\begin{document}

\maketitle

\begin{abstract}
In this paper, we investigate the Unruh radiation in the Bohmian field theory on a shape dynamics background setting. Since metric and metric momentum are real quantities, the integral kernel to invert the Lichnerowicz-York equation for first order deviations due to existence of matter terms turns out to be real. This fact makes the interaction Hamiltonian real. On the other hand, the only contribution to guarantee the existence of Unruh radiation has to come from the imaginary part of the temporal part of the wave functional. We have proved the existence of Unruh radiation in this setting. It is also important that we have found the Unruh radiation via an Unruh-DeWitt detector in a theory where there is no Lorentz symmetry and no conventional space-time structure.
\end{abstract}

\section{Introduction}

Shape dynamics (SD) \cite{sd-1,sd-found-1,sd-found-2,link1,link2} (see ref. \cite{tutorial} for a review) is a theory of gravitation that is based on 3-conformal geometries where time is absent, based on Julian Barbour's interpretation of the Mach's principle \cite{mp}. Although SD agrees with general relativity locally (hence passing all the tests in the  low curvature regime) it may have global differences: for example the spherically symmetric vacuum solution is not the Schwarzschild black hole but rather a wormhole \cite{birkhoff}. Moreover space-time is an emergent phenomenon in SD and it exactly ceases to emerge at the event horizon of the spherically symmetric vacuum solution of SD \cite{birkhoff}. This fact may be related to the firewall paradox and its solution. In ref. \cite{birkhoff} this was also speculated in the context of ER=EPR \cite{er-epr} scenario. See ref. \cite{rot-bh-sd} for more speculation on SD resolution of the firewall paradox.

Bohmian mechanics (BM) \cite{bm-found-1,bm-found-2,bm-found-3} (see ref. \cite{bm} for a review) is an observer-free interpretation of quantum mechanics. There, the quantum particles have definite positions whereas the whole system is always $|\psi|^2$ distributed. Hence Born's statistical interpretation of quantum mechanics is respected. On the other hand Bohmian \emph{field theory} (BFT) \cite{bft1,bft2,bft3,bft4,bft5,bft6,bft7,bft8,bft9} is an attempt to give quantum field theory a Bohmian interpretation.

In this paper we give a derivation for the existence of the Unruh effect \cite{unruh} from the perspective of Bohmian field theory. For this purpose we use a 3-dimensional shape dynamics theory on top of which there is a massive scalar Bohmian field. In order to quantify the existence of Unruh effect, we use the jump rates \emph{i.e.} transition rate explained in Section~\ref{sec:bft} from Bohmian field theory. For this purpose we summarize SD and BFT in Sections \ref{sec:sd} and \ref{sec:bft}. We give a description of scalar fields on shape dynamics background following \cite{tim} in Section~\ref{sec:fieldsd}. In Section~\ref{sec:bftonsd} we give details on how to put Bohmian fields on a shape dynamics background and show the existence of the Unruh effect in this setting.

\section{A Brief Account of Shape Dynamics\label{sec:sd}}

In this part we give a short review of SD which is originally a theory on compact spaces \cite{sd-found-1,sd-found-2}. Its configuration space is the same as that of general relativity \cite{link1,link2}. In SD, the basic ontology is conformal 3-geometries. Hence the theory is symmetric under (volume preserving) conformal transformations. If the reader is interested in the case of asymptotically flat space, Ref. \cite{sd-open} is a good source. However, in our approach we will closely follow Ref. \cite{tim}.

Let $\Sigma$, $N$ and $\xi^a$ respectively be a compact Cauchy surface, a lapse function and a shift vector field. In the ADM (Arnowitt, Deser, and Misner) formalism \cite{adm} the line element of GR is written as follows (in the mostly plus signature):

\begin{equation}
	ds^2 = (g_{ab}\xi^a\xi^b-N^2)dt^2 + 2 g_{ab}\xi^a dx^b dt + g_{ab}dx^a dx^b.
\end{equation}

It is found that $N,\xi^a$ turn out to be Lagrange multipliers of the following constraints:

\begin{align}
    S(N) &= \int_\Sigma d^3x N \Big( \frac{\pi^{ab}(g_{ac}g_{bd}-g_{ab}g_{cd}/2)\pi^{cd}}{\sqrt{g}}\nonumber\\
    &\quad - (R-2\Lambda)\sqrt{g}+\text{matter terms}\Big),\label{eq:Hconst}\\
	H(\xi) &= \int_\Sigma d^3x (\pi^{ab} \mathcal{L}_\xi g_{ab} + \pi^A \mathcal{L}_\xi \phi_A)\label{eq:diffeoconst},
\end{align}

where $\phi_A$ and $\pi^A$ stand for matter fields and their conjugate momenta. Time evolution is generated by $S(N) + H(\xi)$. In the Constant Mean Extrinsic Curvature (CMC) gauge fixing of GR ($\pi - \langle \pi \rangle \sqrt{g} = 0$), there is a conformal factor $\Omega_0$ that solves the Lichnerowicz-York (LY) equation:

\begin{align}
	8\nabla^2 \Omega &= R\Omega + \left( \frac{\langle \pi \rangle ^2}{6} - 2\Lambda \right) \Omega^5 - \frac{1}{g}\Omega^{-7}\sigma^{ab} \sigma_{ab}\nonumber\\ 
    &\quad + \text{matter terms},\label{eq:LY}
\end{align}

where $\sigma^{ab} \equiv \pi^{ab} - \pi g^{ab}/3$ is the trace-free part of the conjugate momentum of the metric. It turns out that the York Hamiltonian, $\int_\Sigma d^3x \sqrt{g} \Omega_0^6$, generates time evolution in York time, $\tau = 2\langle \pi \rangle /3$ where $\pi = \pi^{ab}g_{ab}$ and $\langle \pi \rangle = \int_\Sigma \pi / \int_\Sigma \sqrt{g}$. The smeared gauge-fixging condition

\begin{equation}
	Q(\rho) = \int_\Sigma d^3x \rho \left( \pi - \frac{3}{2}\tau \sqrt{g}\right)\label{eq:Qrho}
\end{equation}

works ``as a generator of spatial conformal transformation for the metric and the trace-free part of the metric momenta and forms a closed constraint algebra with the spatial diffeomorphism generator" \cite{tim}. Therefore the gauge-fixing condition (\ref{eq:Qrho}) can be used to map $\pi \rightarrow \frac{3}{2}\tau \sqrt{g}$ and from now on $\tau$ becomes an ``abstract evolution parameter" \cite{tim} and the total volume of space, $V = \int_\Sigma d^3x \sqrt{g}$, is a pure gauge entity. This new theory is called as shape dynamics with the Hamiltonian

\begin{equation}
	H_{SD} = \int_\Sigma d^3x \sqrt{g}\Omega_0^6[g_{ab},\pi^{ab}\rightarrow \sigma^{ab} + \frac{3}{2}\tau g^{ab}\sqrt{g},\phi_A,\pi^A].
\end{equation}

For the relation between SD and GR readers may see \cite{link1,link2}, where the linking theory approach is used.

\section{Classical massive scalar field on a shape dynamics background\label{sec:fieldsd}}

Let us begin with a usual scalar field on a shape dynamics background. The steps are already sketched in Ref. \cite{tim} and we will follow it closely in this regard. We suppose two properties of the setting: 1) perturbation theory is valid and 2) we look for description on short time intervals. First property is required to use the perturbation theory and the second one is needed to disregard back-reaction of the field on space.

We assume that the Hamiltonian, $H_{SD}$, is even in field quantities, and accepts a perturbative expansion $H_{SD} = H_0 + \varepsilon H_1 + \mathcal{O}(\varepsilon^2)$. Let $\phi$ and $\varpi$ be the massive scalar field and its conjugate field momentum and $\Omega = \Omega_0 + \varepsilon \Omega_1 + \mathcal{O}(\varepsilon^2)$ be the solution of the LY equation~(\ref{eq:LY}), where $\Omega_0$ solves the source-free LY equation. We expand the LY equation as a perturbative series in $\varepsilon$ and the result to first order in $\varepsilon$ is the following:

\begin{equation}
	\Omega_1 = \int_\Sigma d^3y K(x,y)\frac{H^{quad}_{matt}}{\Omega_0\sqrt{g}}(y),
\end{equation}

where $H^{quad}_{matt}$ is quadratic in the field $\phi$ and its momenta and $K(x,y)$ is the Green's function given by:

\begin{equation}
	K(x,y) = \Big[8\nabla^2 - \Big(R + 5 (\frac{3}{8}\tau^2-2\Lambda)\Omega_0^4 +\frac{\sigma^{ab}\sigma_{ab}}{g}\Omega_0^{-8} \Big)\Big]^{-1},\label{eq:K}
\end{equation}

where we used $\tau = 2\langle \pi \rangle /3$. For a massive scalar field we have:

\begin{equation}
	H^{quad}_{matt} = \frac{1}{2}\left(\frac{\varpi^2}{\sqrt{g}} + (\Omega_0^4 g^{ab}\partial_a\phi \partial_b\phi + m^2\phi^2 )\sqrt{g}\right).\label{eq:Hquad}
\end{equation}

Moreover, the first order Hamiltonian, $H_1$, is given as follows:

\begin{equation}
	H_1 = 6 \int d^3x\; d^3y \sqrt{g}(x) \Omega_0^5(x)K(x,y)H^{quad}_{matt}(y).\label{eq:H1}
\end{equation}

Since the lapse function $N(x)$ is the Lagrange multiplier of the Hamiltonian, we obtain:

\begin{equation}
	N(y) = 6 \int_\Sigma d^3x\ \sqrt{g}(x) \Omega_0^5(x)K(x,y).
\end{equation}

\section{Uniformly accelerating observers and shape dynamics variables\label{sec:accobs}}

From now on, we will consider a constantly accelerating observer. The four dimensional Rindler metric for the observer in relativity theory is the following:

\begin{equation}
	ds^2 = -e^{2a\xi} d\lambda^2 + e^{2a\xi} d\xi^2 + dy^2 + dz^2.
\end{equation}

From the perspective of ADM formalism, the shift vector field vanishes, Lapse function equals $N=e^{a\xi}$ and the 3-metric is given by:

\begin{equation}
	g_{ab} = \text{diag}(e^{2a\xi},1,1).
\end{equation}

The spatial compact manifold we consider for SD is a three torus $\Sigma$ with side lengths $L_x,L_y$ and $L_z$ and with volume $V = \prod_i L_i$. The coordinate change to required for moving to $\Sigma$ with Cartesian coordinates is $e^{a\xi} d\xi \rightarrow dx$. The setup is parity symmetric around $x = 0$ \cite{parity-horizon}. This is noted as a \emph{parity horizon} in Ref. \cite{parity-horizon}. (See Ref. \cite{parity-horizon} to see other examples of parity horizons, \emph{e.g.} in SD black holes) In short, $x$ ranges from $0$ to $L_x$, and $\xi$ ranges from $-\infty$ to $L_\xi=\ln(aL_x)/a$.

For Rindler space the momentum conjugate to the three dimensional spatial metric, namely $\pi^{ab}$, vanishes. The Hamiltonian constraint (\ref{eq:Hconst}) is automatically satisfied as well as the diffeomorphism constraint (\ref{eq:diffeoconst}) because the lapse function is zero. The first order LY equation (\ref{eq:LY}) is $\nabla^2 \Omega_0 = 0$. The solution we choose is $\Omega_0 = 1$.

It is quite easy to solve equation~(\ref{eq:K}) for $K(x,y)$ under these circumstances. It is proportional to the Green's function for the Laplacian on the 3-torus. However there is a subtlety in tori in terms of the Green's function equation \cite{green}. For example, the equation to be satisfied by the Green's function on 3-torus with volume $V$ is:

\begin{equation}
	\nabla^2_x G(\vec{x},\vec{y}) = \delta(\vec{x}-\vec{y}) - \frac{1}{V},
\end{equation}

where the last terms is needed for consistency when the equation is integrated and Gauss' theorem is used. In our case, the function of interest is found to be:

\begin{equation}
	K(x,y) = \frac{1}{8|\vec{x}-\vec{y}|} - \frac{|\vec{x}-\vec{y}|^2}{48 V}.
\end{equation}

where the first term may be replaced by its principle value to avoid a singularity at $\vec{x}=\vec{y}$. This solution is valid in Cartesian coordinates in $\Sigma$. Moreover, this is valid when the average value of $\sqrt{|g|} H^{quad}_{matt}$ over space is zero. Otherwise the Laplace equation is not invertible on compact spaces without boundary.

To conclude this section, note that the reduced configuration space of SD and GR are the same \cite{link1,link2}. The constraint for the Constant Mean Extrinsic Curvature (CMC) gauge is satisfied, since $\pi^{ab} = 0$. However we chose to work with SD variables. Our approach should reproduce the same result when viewed from a GR perspective.

\section{A Brief Account of Bohmian Field Theory\label{sec:bft}}

First of all, in order to make a smooth connection to Bohmian Field Theory (BFT), let us give information on what Bohmian Mechanics (BM) \cite{bm,bm-found-1,bm-found-2,bm-found-3} is. BM is a realistic interpretation of Quantum Mechanics (QM). By QM we mean its Copenhagen interpretation. Due to Heisenberg uncertainty in position and momentum ($\Delta x \Delta p \geq \hbar/2$) of QM, it is impossible to know the precise location and momentum of a particle at an instant of time. However in BM, the exact locations particles are in a way the hidden variables that cannot be accessed by macroscopic objects. In BM, the particles are guided by the wave-function of the system and the evolution is deterministic. All interactions such as electromagnetic interactions happen at the wave-function level. This has caused some apparent ``paradoxes" such as the problem of surreal trajectories in BM (See Ref. \cite{surreal} for detailed information). The problem of surreal trajectories, however, is simply solved by accepting that the interaction between particles occur at the wave-function level.

BM poses exact location for each particle in a system and the Heisenberg uncertainty relation emerges as a result of macroscopic observer's inadequacy of knowing the exact position of a particle. Since particle positions can be known in principle, it is also interesting that BM lets one know through which slit the particle passed through in the famous double slit experiment. See Ref. \cite{bm-visual} for visual illustrations of the double slit experiment in BM.

Suppose there are $N$ particles in a BM system. The wave-function is governed by the usual Schrödinger equation of QM \cite{bm}:

\begin{equation}
	i\hbar \frac{\partial \psi(\vec{q},t)}{\partial t} = -\sum_{k=1}^N \frac{\hbar^2}{2m_k}\nabla^2_k \psi(\vec{q},t) + V(\vec{q})\psi(\vec{q},t),
\end{equation}

where $\vec{q} = (\vec{q}_1,\vec{q}_2,\cdots,\vec{q}_N)$ is the vector that is an element of the configuration space. On the other hand, there is another equation that specifically governs the time evolution of the position of each particle \cite{bm}:

\begin{equation}
	\frac{d \vec{Q}_k}{dt} =: \vec{v}_k^\psi(\vec{Q}) = \frac{\hbar}{m_k} \frac{\Im(\psi^*\nabla_k\psi)}{\psi^*\psi}(\vec{Q}_1,\vec{Q}_2,\cdots,\vec{Q}_n),
\end{equation}

where $\vec{Q} = (\vec{Q}_1,\vec{Q}_2,\cdots,\vec{Q}_n)$ and $\vec{Q}_k \in \mathbb{R}^3$ is the position of the $k^{\text{th}}$ particle and $\Im(\cdot)$ stand for the imaginary part of the term inside the parenthesis. Because the equation that governs the time evolution of the position of particles is first order in time, particle paths do not intersect.

\begin{figure}
	\begin{center}
		\includegraphics{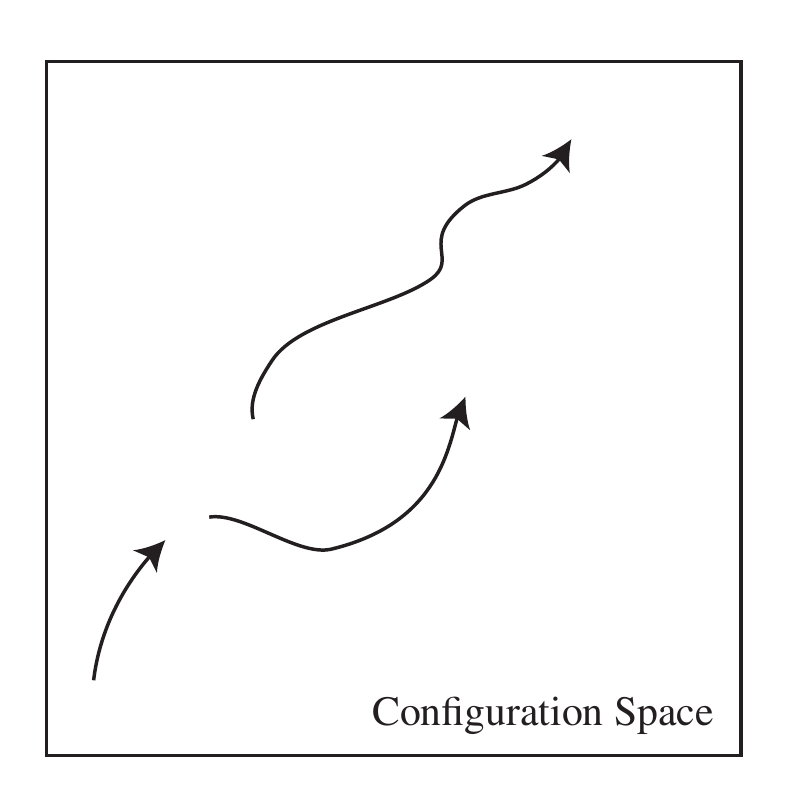}
	\end{center}
    \caption{An illustrative example of a piecewise curve in the configuration space of BFT. Jumps occur when particles are emitted or absorbed.}
    \label{fig:bft-cs}
\end{figure}

So far, this has been BM. BFT is a field theoretical generalization of BM. It is expressed as a stochastic field theory \cite{bft2}. The path that the system follows in the configuration space ($Q$) is piece-wise continuous where jumps occur when particles are emitted or absorbed \cite{bft2}. See Figure~\ref{fig:bft-cs}. In this perspective, it is distinct from BM where path that is followed by the system in the configuration space should be continuous. Moreover, BFT lacks Lorentz symmetry that is inherent in Quantum Field Theory (QFT). BFT uses the Newtonian absolute time as QM does, unlike QFT. Because in SD, there is no 4D refoliation invariance of GR and hence there is no relativity of simultaneity, BFT is more amenable to an SD description rather than the usual QFT.

Let the total Hamiltonian of the system be written as $H=H_0 + H_I$ where $H_0$ is the free Hamiltonian and $H_I$ is the interaction Hamiltonian. In the continuous part of the path in $Q$ the system evolves according to the following formula \cite{bft2}:

\begin{equation}
	\frac{dQ_t}{dt} = \Re \frac{\Psi^*_t(Q_t)(\dot{\hat{q}}\Psi_t)(Q_t)}{|\Psi_t(Q_t)|^2},\label{eq:Qt}
\end{equation}

where $\Re(\cdot)$ stands for the real part of the expression inside the parenthesis and 

\begin{equation}
	\dot{\hat{q}}=\frac{d}{dt} e^{iH_0 t}\hat{q}e^{-iH_0 t}=i[H_0,\hat{q}].
\end{equation}

Here the wave-function $\Psi_t(q)$ evolves according to the Schrödinger equation. When jumps occur, $\sigma^\Psi(dq|q')$ is the jump rate in $dt$ seconds from a point $q'\in Q$ to a volume $dq$ around a point $q \in Q$ \cite{bft2}:

\begin{equation}
	\sigma^\Psi(dq|q') = 2\frac{[\Im \Psi^*(q)\bra{q}H_I\ket{q'}\Psi(q')]^+}{|\Psi(q')|^2},\label{eq:jumprate}
\end{equation}

where $x^+$ expresses the positive part of $x$, \emph{i.e.} $x^+ \equiv \max(x,0)$.

\section{Massive scalar Bohmian field on a shape dynamics background and the Unruh effect\label{sec:bftonsd}}

Unruh effect \cite{unruh}, is the observance of radiation in the  Minkowski vacuum state of rectilinear observers from the perspective of accelerated observers. In its derivation, Lorentz symmetry of space-time and the field are used. It is therefore an open question whether this effect is existent or not in Bohmian field theory on a shape dynamics background.

We consider fields in the Schrödinger wavefunctional picture. Since the Hamiltonian is independent of time, we have:

\begin{equation}
	 \int_\Sigma H_{matt}^{quad} \Psi =  \int_\Sigma \frac{1}{2}\left(-\frac{\delta^2}{\delta\phi^2} + |\nabla\phi|^2 + m^2\phi^2\right)\Psi = 0.\label{eq:timeindepsch}
\end{equation}

Equation (\ref{eq:timeindepsch}) is the time independent Schrödinger equation with energy $E=0$. This is because the integral $\int_\Sigma H_{matt}^{quad}$ should vanish for a well defined physical system on a compact space without boundary as the three-torus we work with. This result is meaningful from a shape dynamics perspective as well. In shape dynamics the total energy of the system should vanish for a reparametrization invariant theory.

We wish to use an Unruh-DeWitt detector to infer the existence Unruh radiation. In short, an Unruh-DeWitt detector is a two-level quantum system that is idealized as a point-like detector \cite{unruh-dewitt-phd}. These detectors are coupled to the Hamiltonian by adding a term $-\lambda \chi \theta \phi$ in QFT where $\lambda$ is a small coupling constant and $\theta$ ``is the operator of the detector's monopole moment" \cite{unruh-dewitt} and $\chi$ is a smooth switching function \cite{unruh-dewitt-phd}. Here $\chi$ denotes the activity of the Unruh-DeWitt detector. If it is nonzero for more time, the detector works for more time. Since there is no previous work on Unruh-DeWitt detectors in the context of BFT we add a real valued function of these variables to the Hamiltonian: $f(\lambda,\chi,\theta,\phi)$. This would not change our final result about the existence of Unruh effect in SD from a BFT perspective. This is because the expectation value of the interaction Hamiltonian will be a multiplicative term to the part that shows the non-zero Unruh radiation. After the inclusion of the detector as an unknown real valued function of the detector variables, the quadratic Hamiltonian becomes:

\begin{equation}
	H_{matt}^{quad} = \frac{1}{2}\left(-\frac{\delta^2}{\delta\phi^2} + |\nabla\phi|^2 + m^2\phi^2\right) +f(\lambda,\chi,\theta,\phi).
\end{equation}

Hence, the true time-independent Schrödinger equation is:

\begin{equation}
	\int_\Sigma H_{matt}^{quad} \Psi =
    	\int_\Sigma \left[\frac{1}{2}\left(-\frac{\delta^2}{\delta\phi^2} + |\nabla\phi|^2 + m^2\phi^2\right) + f(\lambda,\chi,\theta,\phi) \right]\Psi = 0.
\end{equation}

Since the detector is a two level system with energies $E_0,-E_1$ (why there is no minus sign in front of $E_0$ will be clear below) the uncoupled Hamiltonian can yield two energy states with $-E_0,E_1$. This is because the total energy should vanish. On the other hand, we denote $E_0$ as the vacuum energy hence we equate it to zero: $E_0 = 0$. The fact that field and the Unruh-DeWitt detector have opposite energy levels means that if the detector is at energy level $-E$, the field should be at energy $E$. The interaction Hamiltonian $H_I$ on the other hand is the following:

\begin{equation}
	H_I = 6\int_{\Sigma\times\Sigma} K(x,y) H_{matt}^{quad}(y).
\end{equation}

The jump rate for BFT (\ref{eq:jumprate}) is then,

\begin{equation}
	\sigma^\Psi = 2 \frac{[\Im \Psi^*(q,t) H_I \Psi(q',t)]^+}{|\Psi(q',t)|^2}.
\end{equation}

In order to calculate the jump rate, we should not forget the $e^{-iE_i t}$ time dependence of the wave functionals (The field can have energy levels $E = E_1$ and $E = 0$).

\begin{align}
	\sigma^\Psi &= 2 \frac{[\Im e^{-i(E_1-E_0)t}\Psi^*(q) H_I \Psi(q')]^+}{|\Psi(q')|^2}.\\
    \intertext{We use $E_0 = 0$ and obtain the nonzero transition rate:}
    \sigma^\Psi &= 2 \frac{\Psi(q) H_I \Psi(q')}{|\Psi(q')|^2} \sin^+(-E_1 t).
\end{align}

\begin{figure}
	\begin{center}
		\includegraphics{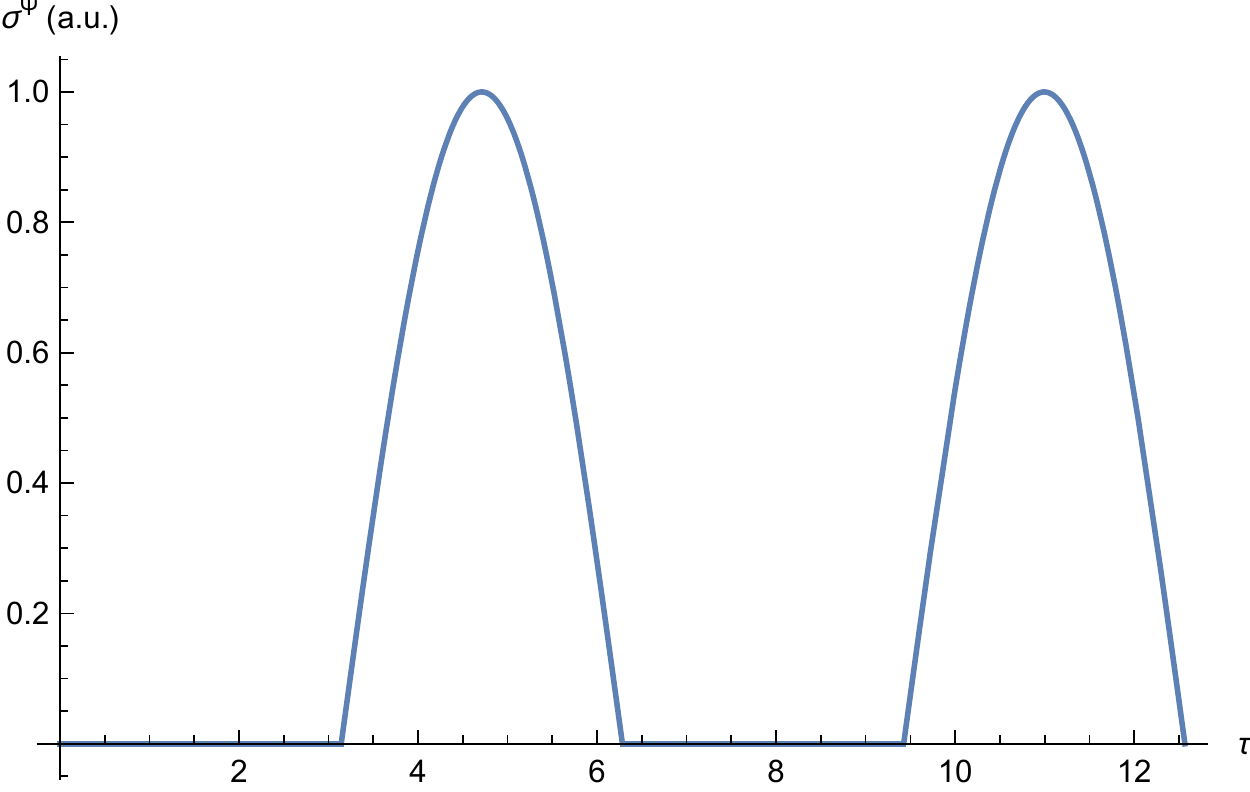}
	\end{center}
    \caption{The jump rate from vacuum to one particle state of the field. The time is rescaled such that $\tau = E_1 t$. The vertical axis is $\sigma^\Psi$ in arbitrary units.}
    \label{fig:sigmaPsi}
\end{figure}

We have shown that there is a non-vanishing transition rate from vacuum to a state with one particle. See Figure~\ref{fig:sigmaPsi}. The term added to the Hamiltonian in order to include the Unruh-DeWitt detector includes a function of detector variables. This does not annul the final result that there is non-zero Unruh radiation: It only effects the spatial distribution of Unruh radiation. In this approach the number of particles is directly related to the number of available states of the Unruh-DeWitt detector. We have been unable to exactly determine the particle production rate, however we have at least shown that there is Unruh radiation detected by Unruh-DeWitt detectors for a Bohmian field theory on a shape dynamics background.

\section{Conclusion}

In this paper we investigated Bohmian field theory on a shape dynamics background. Our aim has been to verify the existence of Unruh radiation for accelerated observers. However because the integral kernel to calculate the first order correction to the solution of the Lichnerowicz-York equation is real, it turned out that the interaction Hamiltonian, $H_I$, is real as well. Hence the required imaginary part to determine a nonzero particle production rate has come from the temporal part of the wave functional.

The existence of Unruh radiation is well known in the standard physics literature. Its existence is strongly tied to Lorentz symmetry of the Minkowski space-time in the usual approach \cite{unruh}. It may be therefore due to lack of Lorentz symmetry both in Bohmian field theory and shape dynamics that one may think there is no Unruh radiation in this setting. However we have proved that there is a non-vanishing transition rate from the vacuum state to an excited state of the Bohmian field.

On the other hand, it is known that Unruh radiation and Hawking radiation are strongly connected with each other. Unruh modes around the black hole's event horizon are outgoing Hawking modes when back-reaction and the angular momentum barrier around the black hole are neglected. Our result hence also motivates further studies on the nature of Hawking radiation from shape dynamics black holes \cite{birkhoff,rot-bh-sd} which are wormholes.

\section{Acknowledgements}

Authors are grateful to Tim Koslowski, Mikhail Sheftel, Edward Anderson, Cihan Pazarbaşı and Bayram Tekin for useful discussions. Authors also acknowledge the proactive criticism of an anonymous referee. F.S.D. is supported by TUBİTAK 2211 Scholarship. The research of F.S.D. and M.A. is partly supported by the research grant
from Boğaziçi University Scientific Research Fund (BAP),
research project No. 11643.

\bibliographystyle{plain}
\bibliography{refs}

\end{document}